%% file: mainArxv.tex
\documentclass[journal,hidelinks]{IEEEtran}
\IEEEoverridecommandlockouts
\usepackage{cite}
\usepackage{amsmath,amssymb,amsfonts, amsmath, dsfont}
\usepackage{graphicx}
\usepackage{textcomp}
\usepackage{xcolor}
\usepackage{multirow}
\usepackage{subcaption}
\usepackage{cite}
\usepackage{textcomp}
\usepackage{eurosym}
\usepackage{dirtytalk}
\usepackage{textcomp}
\usepackage{makecell}
\usepackage{siunitx}

\usepackage[utf8]{inputenc}
\usepackage{booktabs}   
\usepackage{tcolorbox}  
\usepackage{amsmath}    
\usepackage{enumitem}
\usepackage{orcidlink}

\usepackage{tikz}
\usepackage{xcolor, colortbl}
\usetikzlibrary{decorations.pathreplacing}
\usetikzlibrary{arrows,shapes,positioning}
\usetikzlibrary{patterns}
\usetikzlibrary{arrows.meta}
\usepackage{pgfplots}
\pgfplotsset{compat=newest}
\pgfplotsset{plot coordinates/math parser=false}
\pgfkeys{/pgf/number format/.cd,1000 sep={\,}}
\usetikzlibrary{plotmarks}

\usepgfplotslibrary{fillbetween}

\usepackage{pgfplotstable}
\usepgfplotslibrary{groupplots}
\pgfplotsset{compat=1.18}

\newlength\matlabfigurewidth
\newcolumntype{L}[1]{>{\raggedright\let\newline\\\arraybackslash\hspace{0pt}}m{#1}}
\newcolumntype{C}[1]{>{\centering\let\newline\\\arraybackslash\hspace{0pt}}m{#1}}
\newcolumntype{R}[1]{>{\raggedleft\let\newline\\\arraybackslash\hspace{0pt}}m{#1}}

\usepackage{algorithm}
\usepackage{algpseudocode}
\usepackage{multirow}
\usepackage{booktabs}
\usepackage{hhline}
\usepackage{array}
\usepackage{tabularx}

\usepackage{balance}
\usepackage{flushend}

\newcolumntype{J}{>{\centering\arraybackslash}p{0.97cm}}
\newcolumntype{S}{>{\centering\arraybackslash}p{0.80cm}}
\newcolumntype{F}{>{\centering\arraybackslash}p{0.65cm}}

\newcolumntype{C}{>{\centering\arraybackslash}X}

\renewcommand{\tabularxcolumn}[1]{m{#1}}

\begin{document}
%
\title{
A Two-Stage Optimization Framework for Validating Electric Vehicle Charging Infrastructure under Grid Constraints
}

\author{\IEEEauthorblockN{Biswarup Mukherjee, Member IEEE}\\
\IEEEauthorblockA{\textit{{Technical University of Munich}}\\
}
}

\author{Biswarup Mukherjee$^{\orcidlink{0000-0001-7553-4339}}$,~\IEEEmembership{Member,~IEEE}%
\thanks{Corresponding author: Biswarup Mukherjee (E-mail: bismuk@ieee.org).}%
\thanks{}%
}


\maketitle


\begin{abstract}
This paper develops a sequential planning and operational-validation framework to assess post-deployment performance of electric vehicle (EV) charging infrastructure with mixed charger technologies for large fleets under distribution-grid constraints.
The first stage constructs three infrastructure configurations: a cost-oriented design, an investment-constrained integrated design reference, and a same-hardware spatial redistribution. 
In the second stage, each fixed configuration is re-optimized under a common charging-service objective, with selected cases further evaluated through local downward-flexibility redispatch. 
Results on the CIGRE medium-voltage (MV) network show that infrastructure requirements increase with EV fleet size and battery capacity, while spatial redistribution can degrade charging service and shift nodal voltage and loading patterns despite unchanged hardware and investment.
Re-optimizing infrastructure within the first stage investment limit produces only marginal, case-dependent changes in post-planning service. 
Flexibility validation indicates that realizable flexibility through redispatch depends on deployment design and service-degradation allowance, with larger allowances enabling greater delivery in constrained cases.
The framework reveals how infrastructure investment and siting affect post-deployment grid performance, and operational flexibility.
\end{abstract}


\begin{IEEEkeywords}
Distribution networks; Electric vehicles; Optimal planning; Optimal scheduling.
\end{IEEEkeywords}

\section*{Nomenclature}
\noindent\textbf{Abbreviations} \\
\begin{tabular}{p{1cm} p{6.8cm}}
AC & Alternating current \\
CAPEX & Capital expenditure \\
DSO & Distribution system operator \\
EV & Electric vehicle \\
kVA & kilovolt-ampere \\
kW & kilowatt \\
kVAr & kilovar \\
kWh & kilowatt-hour \\
MIP & Mixed-integer programming \\
MILP & Mixed-integer linear programming\\
MV & Medium voltage \\
OPF & Optimal power flow \\
p.u. & Per unit \\
SOC & State-of-charge \\
\end{tabular}

\smallskip

\noindent\textbf{Sets and Indices} \\
\smallskip
\begin{tabular}{p{1cm} p{7.4cm}}
$\mathcal{N}, \mathcal{M}$ & Set of network nodes, indexed by $n$ and $m$\\
$\mathcal{V}$ & Set of EVs, indexed by $i$ \\
$\mathcal{T}$ & Set of time steps, indexed by $t$ \\
$\mathcal{J}$ & Set of charger types, indexed by $j$ \\
$\mathcal{F}$
& Set of active flexibility-request intervals \\
\end{tabular}

\smallskip
\noindent\textbf{Parameters} \\
\smallskip
\begin{tabular}{p{1cm} p{7.4cm}}
$C_j,\;P_j$ & Charger installation cost in € and rated power in kW for type $j$ \\
$\alpha_j,\;\eta$ & Ports per charger unit and charging efficiency in \% \\
$E_i$ & Battery capacity of EV $i$ in kWh \\
$\bar P^{\mathrm{br}}_{}, \bar I_{}$ & Linear active-power branch limit in kW and ampacity, respectively\\


$V^{lim}$ & Voltage limits $[V^{min}, V^{max}]$ in p.u. \\

$\epsilon,\;\Delta t$
& Weighting parameter and time-step duration in h. \\
$w_1,w_2$
& Weighting factors of the charging-service objective \\
$G_{nm}$ & Nodal conductance between nodes $n$ and $m$ \\
$B_{nm}$ & Nodal susceptance between nodes $n$ and $m$ \\

$SOC_i^{tar}$ & Target SOC for EV $i$ at departure \\
$SOC_i^{\mathrm{init}}$
& Initial SOC of EV $i$ \\

$SOC_i^{\mathrm{min}}$
& Minimum terminal SOC required for 
feasibility \\

$\kappa,\;\lambda^{\mathrm{flex}}_{n,t}$
& Loading threshold for flexibility activation and flexibility remuneration in €/kWh \\



$R^{\mathrm{flex}}_{n,t}$
& DSO downward flexibility request at node $n$ and time $t$ in kW \\
$\bar{P}^{\mathrm{grid}}_n$
& Adopted nodal active-power interface/hosting limit at node $n$ in kW \\

$\Delta^{\mathrm{serv}}$
& Maximum allowable degradation of the reference service objective \\

$P^{\mathrm{EV,ref}}_{n,t}$
& Aggregate reference EV charging demand at node $n$ and time $t$ in kW \\


$J^{\mathrm{serv}}$
& Charging-service performance objective \\

\end{tabular}


\noindent\textbf{Variables} \\
\begin{tabular}{p{1cm} p{6.4cm}}
$N_{n,j}$ & Number of chargers of type $j$ at node $n$ \\
$x^{(\cdot)}_{i,t}$ & Binary: 1 if EV $i$ is connected at $t$ \\
$y^{(\cdot)}_{i,t}$ & Binary: 1 if EV $i$ is charging at $t$ \\
$p_{i,t}, q_{i,t}$ & EV active/reactive power at $t$ in kW/kVAr\\
$P^{\text{net}}_{n,t}$ & Net active power 
in kW \\
$Q^{\text{net}}_{n,t}$ & Net reactive power 
in kVAr\\
$V_{n,t}$  & Voltage magnitude at node $n$ at time $t$ in p.u.\\
$\theta_{n,t}$  & Voltage angle at node $n$ at time $t$ in radians \\
$SOC_{i,t}$ & SOC of EV $i$ at $t$ in \% \\
$f^{\mathrm{flex}}_{n,t}$
& Accepted downward flexibility 
in kW \\
$p^{M}_{i,t}$
& EV power after flexibility redispatch in kW\\
$\mathbf{z}^{M}$
& Stage-2B operational decision vector\\

\end{tabular}

\section{Introduction}
By the end of 2022, 2.7 million public EV charging points were installed globally, with more than 900,000 deployed in that year alone \cite{iea_ev_2024}. While it is difficult to accurately predict future charging needs, projections suggest that by 2030, between 280,000 and 480,000 public charging points will be required, which is more than ten times the current number of approximately 25,000 \cite{UKGov2021}. In 2022, the United States saw the installation of 6,300 fast EV chargers. During the same year, the number of slow chargers in the United States grew by 9\%, while in Europe, the total number of slow chargers increased by 50\% compared to the previous year \cite{iea_ev_2024}. Projections by industry analysts \cite{mckerracher2021electric} indicate a significant surge in the global quantities of light-duty EVs and their corresponding charging plugs by 2035, exceeding 300 million and 175 million, respectively.

Thus, the deployment of EV charging infrastructure will require substantial investment, as stations must account not only for installation costs but also for necessary power grid upgrades and integration requirements \cite{brief2019electric}. To effectively accommodate the charging demand of EVs within distribution networks, DSOs must carefully plan the spatial allocation and technological configuration of charging infrastructure to avoid overloading and violations of grid constraints. In this context, smart charging strategies and coordinated planning approaches are increasingly recognized as essential for managing peak demand, improving grid utilization, and reducing the need for costly network reinforcements \cite{brief2019electric}.

\section{Related work and contributions}
Research on EV charging infrastructure addresses two closely related questions: how charging assets should be sited and sized, and how a fixed infrastructure configuration performs during grid-constrained operation. Planning studies such as \cite{mukherjee2021optimal, encej212080} incorporate network limits, heterogeneous charger technologies, and EV connection flexibility to determine suitable infrastructure configurations. These formulations provide insights for charger siting, sizing, and technology selection, but their primary objective is planning-stage design rather than the ex-post comparison of alternative fixed deployments under a common charging-service criterion. This distinction is important because the spatial concentration and technological composition of charging infrastructure directly influence local grid loading and the operational flexibility available after installation \cite{katarinaIET, knezovic2017active, mukherjee2023optimization, RTE, iea_ev_2024}. 
Consequently, infrastructure quality cannot be assessed from investment cost or planning feasibility alone; the charging service ultimately delivered by the fixed deployment must also be examined.

Complementing infrastructure-planning studies, a large body of research has examined the operational scheduling of EV charging. Deterministic approaches \cite{DOLGUI2024, ISGT2021V2G, CIRED2021, sossan:hal-03756809} 
provide tractable formulations for large-scale coordinated charging, while stochastic methods \cite{8810912, WANG2020119886, SEDDIG2019769, WU201755, 9112248, 8467924} 
explicitly represent uncertainty in vehicle availability and charging demand. Although these developments improve the realism of operational decision-making, 
they primarily address charging-schedule design and provide limited insight into how infrastructure choices translate into realized performance once deployment decisions are fixed.
\textcolor{black}{Within the broader operational literature, two-stage and multi-stage formulations have also been introduced to represent sequential charging decisions and temporal coordination more explicitly \cite{ZHOU2024110194, ZHANG2023100262, wang2020two, 9864612}. 
These formulations better capture operational flexibility and the timing of system operator decisions;
however, their primary emphasis remains on charging scheduling once infrastructure is available, rather than on comparing the operational consequences of alternative infrastructure layouts.}
\textcolor{black}{Planning and operations have also been coupled more explicitly via bi-level, sequential, and integrated modeling approaches \cite{prakash2022bi,AlRousan2026,11039016}. These studies establish that linking infrastructure design with charging operations is already well established. However, they do not explicitly isolate, under a common post-planning operational condition, the separate effects of investment budget, spatial allocation, and subsequent EV charging flexibility.}
\textcolor{black}{Motivated by the gap in previous work, we formulate the study as a controlled post-planning comparison problem: once a cost-oriented infrastructure configuration has been obtained, how effectively does the fixed investment deliver recharging service under distribution-grid constraints? 
Our proposed framework addresses this question by comparing alternative infrastructure decisions after planning stage while controlling for infrastructure expenditure and subsequently examining the operational value of local EV flexibility. 
This structure allows observed differences in charging service and grid performance to be attributed more directly to investment, spatial allocation, or operational coordination.}

\begin{table}[ht!]
\centering
\caption{Comparison of the proposed framework with existing planning and operational studies.}
\label{tab:literature_comparison}

\footnotesize
\setlength{\tabcolsep}{1.2pt}
\renewcommand{\arraystretch}{1.03}

\renewcommand{\tabularxcolumn}[1]{m{#1}}

\begin{tabularx}{\columnwidth}{@{}c|ccC|C|C@{}}

\toprule

\multirow{2}{*}{\textbf{Ref.}}
& \multicolumn{3}{c|}{\textbf{Scope}}
& \multirow{2}{*}{\textbf{Comparison}}
& \multirow{2}{*}{\shortstack{\textbf{EV flex.}\\ \textbf{Option}}}
\\

\cmidrule(lr){2-4}

& \textbf{Plan.}
& \textbf{Grid}
& \textbf{Operation}
& &
\\
\hline

\midrule

\cite{mukherjee2021optimal,encej212080}
& $\checkmark$
& $\checkmark$
& Embedded
& \shortstack{Tech. choice/\\ CAPEX/OPEX}
& \shortstack{Plug /\\unplug}
\\

\cmidrule(lr){1-6}

\cite{SEDDIG2019769,WU201755,9112248}
& --
& Partial
& \shortstack{Stoch. 2-/\\multi-stage}
& \shortstack{Heuristic/\\deterministic}
& \shortstack{Temporal\\load shift}
\\

\cmidrule(lr){1-6}

\cite{prakash2022bi,AlRousan2026}
& $\checkmark$
& $\checkmark$
& \shortstack{Bi-level /\\sequential}
& \shortstack{Uncoord. /\\siting}
& \shortstack{V2G\\scheduling}
\\

\cmidrule(lr){1-6}

\cite{11039016}
& $\checkmark$
& $\checkmark$
& \shortstack{Integrated /\\real-time}
& --
& \shortstack{EVCS \\control}
\\

\midrule

\textbf{This work}
& $\checkmark$
& $\checkmark$
& \textbf{\shortstack{Fixed-plan\\re-optimization}}
& \textbf{O / I / U}
& \textbf{\shortstack{Local flex.\\activation}}
\\

\bottomrule

\end{tabularx}
\end{table}
\begin{figure}[t]
    \centering
    \includegraphics[width=0.98\columnwidth]{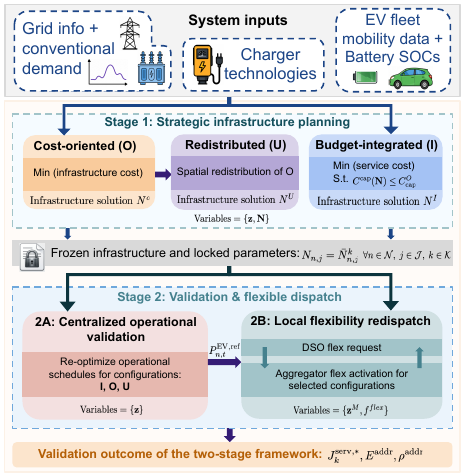}
    \caption{Planning and operational-validation workflow.
    }
    \label{fig:validation_framework}
\end{figure}
\textcolor{black}{Table~\ref{tab:literature_comparison} summarizes the proposed framework with representative planning and operational studies, while Fig.~\ref{fig:validation_framework} illustrates the complete planning--validation workflow. Rather than the generic two-stage architecture itself, the contribution of this work lies in evaluating three deliberately constructed infrastructure alternatives: O, U, and I. 
These configurations represent, respectively, retaining the original cost-oriented layout, redistributing the charger portfolio across eligible grid nodes, and reallocating infrastructure while remaining within the available investment budget.
Their operational performance is assessed using a common 
charging-service formulation with smart-charging schedules, after which selected cases are subjected to local-flexibility request and activation to examine whether further temporal rescheduling of EV charging can alleviate localized grid stress while preserving acceptable charging-service performance and without modifying the installed charging infrastructure.
Here, \say{validation} denotes post-planning operational assessment of fixed infrastructure designs rather than empirical validation against field measurements.}
\textcolor{black}{The main contributions of this paper are summarized as follows:
\begin{enumerate}[
    label={},
    leftmargin=0pt,
    itemindent=0pt,
    itemsep=1pt,
    topsep=2pt,
    parsep=0pt
]
\item \hspace{1.2em}(1)\hspace{0.35em}We develop a structured planning--operation validation framework for large EV fleets that quantifies how a cost-oriented infrastructure design translates into realized charging-service and grid performance after deployment.
\item \hspace{1.2em}(2)\hspace{0.35em}The framework further introduces an investment-constrained integrated design reference that jointly optimizes infrastructure allocation and charging service within the Stage-1A CAPEX limit, thereby isolating service improvements achievable without additional infrastructure expenditure.
\item \hspace{1.2em}(3)\hspace{0.35em}To isolate the effect of siting, a same-hardware spatial-redistribution reference is introduced to emulate deployment under nodal-space limitations.
It preserves the Stage-1A investment and installed hardware while altering only the spatial allocation of the installed chargers across eligible grid nodes.
\item \hspace{1.2em}(4)\hspace{0.35em} An aggregator-based local-flexibility layer is introduced that quantifies the EV-addressable portion of DSO flexibility requests and the realizable response under explicit charging-service degradation limits, without modifying the installed infrastructure.
\end{enumerate}
}

The remainder of the paper is organized as follows. Section~III presents the methodology and underlying assumptions. Section~IV describes the case study and system parameters. Section~V presents the results and discussion, and Section~VI concludes the paper.

%
%

\section{Methodology}\label{sec:methodology}

\textcolor{black}{This section presents the mathematical formulation of the proposed two-stage grid-aware framework for EV charging infrastructure planning and post-planning operational assessment. The problem is inherently coupled because long-term infrastructure decisions directly affect the feasibility and quality of short-term charging operation under distribution-network constraints.
To represent this interaction consistently, each stage of the formulation combines EV charging dynamics, charger availability and capacity, and distribution grid constraints within a common optimization model. Stage~1 determines the charging-infrastructure configuration, including mixed charger technologies, locations, and available ports, while enforcing operational feasibility. Stage~2 fixes the resulting infrastructure and re-optimizes EV charging operation under the same physical constraints for comparative validation and local-flexibility assessment. This common modeling basis allows differences in post-planning performance to be attributed to the infrastructure configuration rather than to changes in the underlying operational constraints.}
%

\subsection{System modeling}

The system model integrates power system constraints with EV charging behavior to ensure both physical realism and grid feasibility. The main components are described below.

\subsubsection{EV charging dynamics and constraints}

The operational behavior of the EV fleet is governed by battery energy evolution and hardware-specific interconnection logic. Let $\mathcal{T} = \{0, 1, \dots, T-1\}$ denote the set of discrete time intervals of duration $\Delta t$, and $\mathcal{V}$ be the set of all EVs. For each vehicle $i \in \mathcal{V}$, the SOC transition is modeled as:
\begin{equation}
\label{eq:soc_dynamics}
SOC_{i,t+1} = SOC_{i,t} + \frac{\eta \, p_{i,t} \, \Delta t}{E_i}, \quad \forall t \in \mathcal{T}
\end{equation}
where $\eta$ and $E_i$ represent the charging efficiency and battery energy capacity, respectively. \textcolor{black}{The initial battery state is fixed in both planning and validation as $SOC_{i,0}=SOC_i^{\mathrm{init}}$, for all $i\in\mathcal V$.}

To characterize the discrete nature of charging states, we introduce binary decision variables. Let $x^{f}_{i,t}, x^{s}_{i,t} \in \{0, 1\}$ denote the physical connection status of vehicle $i$ to a fast or slow charger at time $t$, respectively. Similarly, let $y^{f}_{i,t}, y^{s}_{i,t} \in \{0, 1\}$ represent the active power delivery state. The following logical constraints ensure mutually exclusive and consistent operation:
\begin{align}
\label{eq:logic_plug}
x^{f}_{i,t} + x^{s}_{i,t} &\leq 1, \quad \forall i \in \mathcal{V}, \forall t \in \mathcal{T} \\
\label{eq:logic_charge}
y^{f}_{i,t} \leq x^{f}_{i,t}, \quad & y^{s}_{i,t} \leq x^{s}_{i,t}, \quad \forall i \in \mathcal{V}, \forall t \in \mathcal{T}.
\end{align}
Constraint \eqref{eq:logic_plug} prevents simultaneous connection to multiple charger types, while \eqref{eq:logic_charge} ensures that active charging occurs only when a physical connection is established. 
\textcolor{black}{Finally, the instantaneous charging power $p_{i,t}$ is modeled as a continuous decision variable bounded by the selected charger type:
\begin{equation}\label{eq:powerlimit}
p_{i,t} \leq P^f y^f_{i,t} + P^s y^s_{i,t}, \quad \forall i \in \mathcal{V}, t \in \mathcal{T}.
\end{equation}
This formulation allows flexible (modulated) charging within the rated capacity of the assigned charger, rather than enforcing fixed power levels}.

\subsubsection{Grid-constrained power flow model}
To ensure that the spatial and temporal distribution of EV charging demand does not compromise network integrity, a linearized AC-OPF formulation is adopted. For each node $n \in \mathcal{N}$ at time $t \in \mathcal{T}$, the active and reactive power balances are:
\begin{alignat}{1}
&P^{\text{gen}}_{n,t} - P^{\text{load}}_{n,t} - \sum_{i \in \mathcal{V}_n} p_{i,t} = P^{\text{net}}_{n,t}, \quad \forall n,t \label{eq:p_balance} \\
&Q^{\text{gen}}_{n,t} - Q^{\text{load}}_{n,t} - \sum_{i \in \mathcal{V}_n} q_{i,t} = Q^{\text{net}}_{n,t}, \quad \forall n,t \label{eq:q_balance}
\end{alignat}
where $P^{\text{load}}$ and $Q^{\text{load}}$ represent the existing base demand. The relationship between nodal injections and state variables $(V, \theta)$ is defined as:
\begin{alignat}{1}
&\Delta V_{nm,t} = V_{n,t} - V_{m,t}, \quad \Delta \theta_{nm,t} = \theta_{n,t} - \theta_{m,t}, \label{eq:delta_defs} \\
&P^{\text{net}}_{n,t} = \sum_{m \in \mathcal{N}} \big[ G_{nm} \Delta V_{nm,t} - B_{nm} \Delta \theta_{nm,t} \big], \label{eq:p_flow_compact} \\
&Q^{\text{net}}_{n,t} = \sum_{m \in \mathcal{N}} \big[ -B_{nm} \Delta V_{nm,t} - G_{nm} \Delta \theta_{nm,t} \big], \label{eq:q_flow_compact}
\end{alignat}
where $G_{nm}$ and $B_{nm}$ are the nodal conductance and susceptance, respectively. This formulation captures the linearized relationship between nodal power injections, voltage magnitudes $V$, and phase angles $\theta$. 
\textcolor{black}{Voltage limits are enforced as:
\begin{equation}
V_{\min}\leq V_{n,t}\leq V_{\max},
\qquad
\theta_{\min}\leq\theta_{n,t}\leq\theta_{\max}.
\label{eq:voltage_limits}
\end{equation}}
\textcolor{black}{The network branch-current ratings are represented through linear active-power branch limits. For each branch $(n,m)$, the corresponding active-power transfer and its operating limit are written as:
\begin{equation}
\begin{aligned}
P_{nm,t}^{\mathrm{br}}
&=
S^{\mathrm{base}}
\left(
G_{nm}\Delta V_{nm,t}
-
B_{nm}\Delta\theta_{nm,t}
\right),\\
-\bar P_{nm}^{\mathrm{br}}
&\leq
P_{nm,t}^{\mathrm{br}}
\leq
\bar P_{nm}^{\mathrm{br}},
\qquad
\bar P_{nm}^{\mathrm{br}}
=
\sqrt{3}\,V^{\mathrm{base}}\bar I_{nm}.
\end{aligned}
\label{eq:branch_limit}
\end{equation}
In \eqref{eq:branch_limit}, $\bar I_{nm}$ denotes the branch ampacity. Eq.
\eqref{eq:branch_limit} provides a linear active-power surrogate of the branch thermal limit rather than an explicit current-magnitude constraint. Conventional active- and reactive-power demands are treated as exogenous nodal inputs to the network model, with reactive demand obtained from the corresponding active demand and prescribed nodal power factor. Nodal capacity is represented separately through the nodal active-power interface constraint in \eqref{eq:nodal_power_limit}.}

%

\subsubsection{Modeling the infrastructure capacity and nodal constraints}
The available charging capacity at each node is a direct function of the installed infrastructure. We define $\mathcal{J}$ as the set of available charger technologies (i.e., fast/slow, both single-port and multi-port technologies) and $\mathcal{V}_n \subset \mathcal{V}$ as the subset of EVs assigned to node $n \in \mathcal{N}$. 

Let $N_{n,j} \in \mathbb{Z}_{\geq 0}$ denote the integer number of charger units of type $j \in \mathcal{J}$ deployed at node $n$. The aggregate EV nodal power capacity, $P_n^{\max}$, and the total number of physical charging ports, $K_n$, are formulated as:
\begin{align}
\label{eq:cap_power}
P_n^{\max} &= \sum_{j \in \mathcal{J}} P_j \alpha_j N_{n,j}, \quad \forall n \in \mathcal{N} \\
\label{eq:cap_ports}
K_n &= \sum_{j \in \mathcal{J}} \alpha_j N_{n,j}, \quad \forall n \in \mathcal{N}
\end{align}
where $P_j$ represents the rated power per port for technology $j$, and $\alpha_j$ is the port multiplier (e.g., $\alpha_j = 4$ for multi-port columns, $\alpha_j = 1$ for single-port units). 
Multi-port chargers are represented via aggregate port capacity and nodal power limits, without tracking individual port assignments. This lets EVs connect to any available port of a chosen charger type, while ensuring simultaneous connections and total power stay within physical limits at each node.

The deployed infrastructure defines each node’s operating envelope by limiting instantaneous power delivery and physical occupancy. For each time step $t \in \mathcal{T}$, the following nodal constraints are enforced across the network $\mathcal{N}$:
\begin{align}
\label{eq:nodal_power_limit1}
\sum_{i \in \mathcal{V}_n} p_{i,t} &\leq P_n^{\max}, \quad \forall n \in \mathcal{N}, \forall t \in \mathcal{T} \\
\label{eq:nodal_power_limit}
P^{\text{load}}_{n,t} + \sum_{i \in \mathcal{V}_n} p_{i,t} &\leq \textcolor{black}{\bar P_n^{\mathrm{grid}}}, \quad \forall n \in \mathcal{N}, \forall t \in \mathcal{T}
\\
\label{eq:nodal_port_limit}
\sum_{i \in \mathcal{V}_n} \left(x^f_{i,t} + x^s_{i,t}\right) &\leq K_n, \quad \forall n \in \mathcal{N}, \forall t \in \mathcal{T}
\end{align}
Constraint~\eqref{eq:nodal_power_limit1} caps total EV charging power at node $n$ by the rated hardware capacity. \textcolor{black}{Constraint~\eqref{eq:nodal_power_limit} is the linear grid-interface limit used in the optimization, ensuring that the sum of conventional base load and EV charging demand stays below the adopted nodal active-power capacity.} Constraint~\eqref{eq:nodal_port_limit} bounds the number of simultaneously connected vehicles—via the fast ($x^f_{i,t}$) and slow ($x^s_{i,t}$) binary indicators—by the number of ports $K_n$. \textcolor{black}{This decouples individual port assignment from infrastructure sizing, reducing model complexity while retaining the key constraints on simultaneous connections and aggregate charging power.}

\subsection{The sequential planning--validation framework}

\textcolor{black}{Our proposed methodology uses a sequential planning--validation structure. Stage~1A obtains the cost-oriented infrastructure configuration O, while Stage~1B constructs the investment-constrained integrated reference I subject to the Stage~1A investment limit. A same-hardware spatial control U is subsequently constructed from O. The infrastructure decisions associated with O, I, and U are then fixed in Stage~2A, where only operational decisions are re-optimized under a common charging-service criterion. Stage~2B further evaluates selected fixed configurations through local downward-flexibility activation. 
All optimization stages are formulated as MILPs.}
\textcolor{black}{
For compactness, let
\begin{equation}
\mathbf{N}=\{N_{n,j}\}_{n\in\mathcal{N},\,j\in\mathcal{J}}
\end{equation}
denote the infrastructure decision vector. The operational decision vector is denoted by:
\begin{equation}
\mathbf{z} =
\left(
x^{(\cdot)}_{i,t},
y^{(\cdot)}_{i,t},
p_{i,t},SOC_{i,t},
P^{\mathrm{net}}_{n,t},
Q^{\mathrm{net}}_{n,t},
V_{n,t},\theta_{n,t}
\right),
\label{eq:operational_vector}
\end{equation}
with the corresponding indices omitted for compactness.
For both the integrated planning--operation reference and the fixed-infrastructure operational validation, charging-service quality is evaluated with a common SOC-based objective. 
Let $T_f$ denote the final time index of the scheduling horizon.
The cumulative and terminal SOC deviations are defined as:
\begin{align}
J_{\mathrm{cum}}
&=
\sum_{i\in\mathcal V}\sum_{t\in\mathcal T}
\left(
SOC_i^{\mathrm{tar}}-SOC_{i,t}
\right),
\label{eq:Jcum}\\
J_{\mathrm{term}}
&=
\sum_{i\in\mathcal V}
\left(
SOC_i^{\mathrm{tar}}-SOC_{i,T_f}
\right).
\label{eq:Jterm}
\end{align}
The combined charging-service objective is expressed as:
\begin{equation}
J^{\mathrm{serv}}(\mathbf z)
=
w_1J_{\mathrm{cum}}
+
w_2J_{\mathrm{term}},
\label{eq:service_obj}
\end{equation}
where $w_2\gg w_1$ prioritizes end-of-horizon service satisfaction while retaining a secondary preference for charging progress over the scheduling horizon.}

\subsubsection{Stage 1A - Strategic infrastructure planning}
\textcolor{black}{The primary objective of Stage~1A is infrastructure investment minimization across all grid nodes. This stage determines the spatial allocation, technology mix, and number of chargers to be installed across grid nodes. Infrastructure expenditure and the aggregate charger-utilization measure are defined as:}
\textcolor{black}{
\begin{equation}
C^{\mathrm{cap}}(\mathbf{N})
=
\sum_{n\in\mathcal{N}}
\sum_{j\in\mathcal{J}}
C_j N_{n,j},
\label{eq:capex}
\end{equation}
%
\begin{equation}
U(\mathbf{x})
=
\sum_{n\in\mathcal{N}}
\sum_{i\in\mathcal{V}_n}
\sum_{t\in\mathcal{T}}
\left(x^f_{i,t}+x^s_{i,t}\right).
\label{eq:utilization}
\end{equation}
A generalized Stage-1 objective is:
\begin{equation}
\min_{\mathbf{N},\mathbf{z}}
\quad
C^{\mathrm{cap}}(\mathbf{N})
-
\epsilon U(\mathbf{x}),
\label{eq:stage1_general}
\end{equation}
where $\epsilon\geq0$ controls an optional secondary preference for charger utilization. In all reported experiments in this paper, $\epsilon=0$; hence, the Stage~1A solution minimizes infrastructure expenditure only as:
\begin{equation}
C_{\mathrm{cap}}^{*}
=
\min_{\mathbf{N},\mathbf{z}}
C^{\mathrm{cap}}(\mathbf{N}),
\label{eq:stage1_primary}
\end{equation}
If a cost-optimal solution is available and a secondary tie-break among CAPEX-equivalent solutions is desired, charger utilization can subsequently be maximized while preserving the optimal investment cost:
\begin{equation}
\max_{\mathbf{N},\mathbf{z}}
\quad
U(\mathbf{x}),
\label{eq:stage1_secondary}
\end{equation}
subject to the same Stage-1A feasibility constraints and
$C^{\mathrm{cap}}(\mathbf{N})=C_{\mathrm{cap}}^{*}$. This optional tie-break is useful when multiple infrastructure configurations attain the same minimum CAPEX, as it selects among them according to charger utilization without increasing investment.}

\textcolor{black}{The Stage-1A decision vector contains the infrastructure variables $\mathbf N$ together with the operational variables required to enforce charging feasibility, SOC dynamics, and network operating constraints. For all applicable $n\in\mathcal{N}$, $j\in\mathcal{J}$,
$i\in\mathcal{V}_n$, and $t\in\mathcal{T}$, the Stage-1 variable domains are:
\begin{align}
p_{i,t}&\in\mathbb{R}_{\geq 0},
&
P^{\mathrm{net}}_{n,t}, Q^{\mathrm{net}}_{n,t}&\in\mathbb{R},
\label{eq:continuous_domains}\\
N_{n,j}&\in\mathbb{Z}_{\geq 0},
&
x^{f}_{i,t},x^{s}_{i,t},y^{f}_{i,t},y^{s}_{i,t}
&\in\{0,1\}.
\label{eq:stage1_domains}
\end{align}
with the remaining continuous variables subject to the bounds defined in Section~III-A.
The reported Stage-1 problem in \eqref{eq:stage1_primary} is solved subject to the EV charging constraints in \eqref{eq:soc_dynamics}--\eqref{eq:powerlimit}, the network model in \eqref{eq:p_balance}--\eqref{eq:nodal_port_limit}, and the SOC feasibility conditions:
\begin{align}
0.05
&\leq SOC_{i,t}\leq 1,
&&\forall i\in\mathcal V,\ t\in\mathcal T,
\label{eq:soc_bounds}\\
SOC_{i,T_f}
&\geq SOC_i^{\mathrm{min}},
&&\forall i\in\mathcal V.
\label{eq:soc_min}
\end{align}
The resulting Stage-1A solution is denoted by
$(\mathbf{N}^{O},\mathbf{z}^{O})$, where $\mathbf{N}^{O}$ defines the cost-oriented infrastructure configuration O.}

\subsubsection{Stage 1B - Investment-constrained integrated planning--operation reference}
\textcolor{black}{After obtaining the cost-oriented infrastructure configuration $\mathbf{N}^{O}$ in Stage~1A, we construct a second planning reference to examine whether the same investment limit can support a different infrastructure allocation with improved charging-service performance. This reference, denoted by $I$, jointly re-optimizes the infrastructure and charging operation while preventing any increase in infrastructure expenditure relative to configuration O. 
The Stage~1A investment cost is first evaluated as:
\begin{equation}
C_{\mathrm{cap}}^{O}
=
C^{\mathrm{cap}}(\mathbf N^{O})
\label{eq:O_capex}
\end{equation}
Unlike Stage~1A, the objective of Stage~1B is not to minimize
infrastructure expenditure. Instead, the expenditure $C_{\mathrm{cap}}^{O}$ is treated as an upper investment limit, and both the infrastructure locations and charging operation are jointly re-optimized to obtain the best charging-service performance achievable within that budget. The integrated reference in Stage~1B is therefore formulated as:
\begin{equation}
\min_{\mathbf N,\mathbf z}
\quad
J^{\mathrm{serv}}(\mathbf z)
\label{eq:integrated_reference}
\end{equation}
subject to the charging, SOC, infrastructure-capacity, and network constraints, together with
\begin{equation}
C^{\mathrm{cap}}(\mathbf N)
\leq
C_{\mathrm{cap}}^{O}.
\label{eq:investment_limit}
\end{equation}
The resulting infrastructure solution is denoted by $\mathbf{N}^{I}$. Thus, $O$ and $I$ solve two different planning questions: $O$ seeks the least-cost infrastructure satisfying the Stage~1 minimum feasibility requirements, whereas $I$ seeks the best charging-service performance without exceeding the investment budget established by $O$. Consequently, any service improvement obtained by $I$ is attributable to a different allocation or composition of the available infrastructure solution rather than to additional CAPEX.}

\subsubsection{Stage 2A - Centralized operational validation}

\textcolor{black}{Before operational validation, a third infrastructure configuration, denoted by $U$, is constructed as a spatial redistribution control. Unlike $O$ and $I$, configuration $U$ is not obtained from an additional planning optimization. 
Instead, it preserves the charger-technology portfolio and aggregate infrastructure investment of $\mathbf{N}^{O}$ while redistributing the installed units more uniformly across the eligible charging locations of the distribution network.
The specific construction of $U$ used in the case study is described in Section~IV-C.
Stage 2A then evaluates the three fixed infrastructure configurations:
\begin{equation}
\mathcal{K}=\{O,I,U\}
\end{equation}
under the same centralized charging-service optimization.
For each configuration $k\in\mathcal{K}$, let
$\bar{\mathbf{N}}^{k}$ denote the corresponding infrastructure configuration obtained before operational validation. Accordingly, $\bar{\mathbf{N}}^{O}$ is derived from Stage~1A, $\bar{\mathbf{N}}^{I}$ from the investment-constrained integrated problem in Stage~1B, and $\bar{\mathbf{N}}^{U}$ from the controlled spatial redistribution of $O$.
During Stage 2A, these infrastructure decisions are treated as fixed parameters rather than optimization variables:
\begin{equation}
N_{n,j}=\bar{N}^{k}_{n,j},
\qquad
\forall n\in\mathcal{N},\;
j\in\mathcal{J},\;
k\in\mathcal{K}.
\label{eq:fixed_infrastructure}
\end{equation}
For each $k\in\mathcal{K}$, Stage 2A re-optimizes only the operational variables $\mathbf{z}$ with objective:
\begin{equation}
J_{k}^{\mathrm{serv},*}
=
\min_{\mathbf{z}}
\quad
J^{\mathrm{serv}}(\mathbf{z}),
\label{eq:centralized_validation}
\end{equation}
subject to the charging, SOC, infrastructure-capacity, and network constraints defined above and the fixed-infrastructure condition in \eqref{eq:fixed_infrastructure}.
The corresponding optimal charging schedule is denoted by $\mathbf{p}^{\mathrm{ref},k}$.
\textcolor{black}{Stage~1B and Stage~2A share the same charging-service objective but have different decision spaces.}
Hence, Stage 2A provides a common operational benchmark for comparing the service delivered by $O$, $I$, and $U$ when their infrastructure is held fixed.}

\subsubsection{Stage 2B -- Local-flexibility redispatch validation}

\textcolor{black}{Stage~2B models a local flexibility interaction between the DSO and an EV aggregator after the centralized Stage-2A schedule has been established. 
For a selected infrastructure configuration $k$, the Stage-2A
charging schedule $p_{i,t}^{\mathrm{ref},k}$ is retained as the baseline schedule prior to local-flexibility activation.
Based on the corresponding nodal loading, the DSO identifies node--time pairs at which the reference load exceeds a prescribed fraction $\kappa$ of the nodal power limit and requests downward flexibility. The aggregator then re-optimizes EV charging to respond to these requests while the installed infrastructure remains fixed and charging-service degradation is bounded.
Vehicle-to-grid discharge is not considered; flexibility is provided exclusively through charging reduction or rescheduling. For compactness, the superscript $k$ is omitted below when the reference configuration is unambiguous.}

\textcolor{black}{The aggregate EV demand associated with the Stage-2A reference is computed as:
\begin{equation}
P_{n,t}^{\mathrm{EV,ref}}
=
\sum_{i\in\mathcal{V}_n}
p_{i,t}^{\mathrm{ref}},
\qquad \forall n,t .
\label{eq:ev_reference}
\end{equation}
Using this reference, the DSO downward-flexibility request and the corresponding active request set are defined with:
\begin{equation}
R_{n,t}^{\mathrm{flex}}
=
\left[
P_{n,t}^{\mathrm{load}}
+
P_{n,t}^{\mathrm{EV,ref}}
-
\kappa\bar P_n^{\mathrm{grid}}
\right]^+ ,
\label{eq:flex_request}
\end{equation}
\begin{equation}
\mathcal{F}
=
\left\{
(n,t)\in\mathcal{N}\times\mathcal{T}
:
R_{n,t}^{\mathrm{flex}}>0
\right\}.
\label{eq:flex_set}
\end{equation}
In \eqref{eq:flex_request}, $[a]^+=\max(a,0)$ denotes the positive-part operator, while $\kappa\in(0,1]$ is an exogenous DSO loading-threshold parameter. 
Consequently, a downward-flexibility request
is generated only when the reference nodal load exceeds $\kappa\bar{P}_n^{\mathrm{grid}}$.}

\textcolor{black}{In response, the aggregator determines the Stage-2B redispatched charging schedule $p_{i,t}^{M}$ and accepted downward flexibility $f_{n,t}^{\mathrm{flex}}$. For every $(n,t)\in\mathcal{F}$,
\begin{equation}
0
\leq
f_{n,t}^{\mathrm{flex}}
\leq
R_{n,t}^{\mathrm{flex}},
\label{eq:flex_bound}
\end{equation}
\begin{equation}
f_{n,t}^{\mathrm{flex}}
\leq
P_{n,t}^{\mathrm{EV,ref}}
-
\sum_{i\in\mathcal{V}_n}p_{i,t}^{M},
\label{eq:flex_actual}
\end{equation}
\begin{equation}
J^{\mathrm{serv}}(\mathbf{z}^{M})
\leq
J_{k}^{\mathrm{serv},*}
+
\Delta^{\mathrm{serv}} .
\label{eq:service_guard}
\end{equation}
Equations~\eqref{eq:flex_bound} and
\eqref{eq:flex_actual} ensure that accepted flexibility cannot exceed either the DSO request or the actual reduction in aggregate EV charging relative to the Stage-2A reference. Equation~\eqref{eq:service_guard} limits the deterioration of charging service caused by the redispatch.}

\textcolor{black}{The aggregator therefore solves
%
\begin{equation}
\max_{\mathbf{z}^{M},\,f^{\mathrm{flex}}}
\quad
\sum_{(n,t)\in\mathcal{F}}
\lambda_{n,t}^{\mathrm{flex}}
f_{n,t}^{\mathrm{flex}}\Delta t ,
\label{eq:stage2b_objective}
\end{equation}
subject to Eq.~\eqref{eq:flex_bound}--\eqref{eq:service_guard} and the same charging, SOC, infrastructure-capacity, and grid constraints used in Stage~2A. Thus, Stage~2B changes only the charging operation: the infrastructure configuration remains fixed throughout the flexibility activation.}

\textcolor{black}{The total DSO-requested energy, EV-addressable upper bound, and realized flexibility are defined as:
\begin{align}
E^{\mathrm{req}}
&=
\sum_{(n,t)\in\mathcal{F}}
R_{n,t}^{\mathrm{flex}}\Delta t ,
\label{eq:Ereq}
\\
E^{\mathrm{addr}}
&=
\sum_{(n,t)\in\mathcal{F}}
\min\!
\left\{
R_{n,t}^{\mathrm{flex}},
P_{n,t}^{\mathrm{EV,ref}}
\right\}\Delta t ,
\label{eq:Eaddr}
\\
E^{\mathrm{flex}}
&=
\sum_{(n,t)\in\mathcal{F}}
f_{n,t}^{\mathrm{flex}}\Delta t .
\label{eq:Eflex}
\end{align}
The EV-addressable delivery ratio is defined as
$\rho^{\mathrm{addr}}=E^{\mathrm{flex}}/E^{\mathrm{addr}}$ and is reported as a percentage in 
Section~\ref{sec:results}.
The quantity $E^{\mathrm{addr}}$ is a pre-redispatch upper bound: it reflects the EV charging available at requested node over time, but does not account for the SOC, charging-service, infrastructure, or network constraints encountered during redispatch. $E^{\mathrm{flex}}$ represents the flexibility that remains realizable after all Stage-2B constraints are enforced.}

%

\subsection{Modeling assumptions and scope}
\textcolor{black}{The overall framework is deterministic. EV mobility follows a home--work commute; initial SOC, spatial assignment, and charging needs are generated as in Section~\ref{sec:EV_details}. 
Scenario-based uncertainty can be handled in either stage by replicating operational constraints across scenarios and using an expected- or worst-case objective.
The distribution grid is modeled with the adopted linearized power-flow formulation. Chargers operate at unity power factor, so $q_{i,t}=0$ for all $i,t$. Charging efficiency is constant. EV flexibility comes only from reducing or rescheduling unidirectional charging. Explicit traction-power discharge during driving is not modeled.
In centralized validation, charging is coordinated system-wide. The local-flexibility extension models an aggregator responding to DSO flexibility requests under exogenous remuneration; strategic bidding, and detailed market clearing are excluded.}

\section{Case study setup and system parameters}
\textcolor{black}{The proposed framework is evaluated on the European CIGRE MV distribution network \cite{cigref,strunz2005cigre} over a 24-hour planning and scheduling horizon. Our case study is designed to separate the effects of infrastructure planning, spatial allocation, and operational coordination. Configurations $O$, $I$, and $U$ are considered for comparative analysis. Each configuration is evaluated under fixed-infrastructure centralized operation in Stage 2A, while selected cases are additionally assessed through the local-flexibility formulation in Stage 2B. Unless otherwise stated, identical grid, mobility, battery, and charging parameters are used across all comparisons.}
\subsection{Grid architecture and charging infrastructure}
\textcolor{black}{The 14-bus CIGRE MV benchmark network operates at 20~kV.
Node~0 is used as the slack/reference node. Voltage magnitudes are constrained within $0.95$--$1.05$~p.u., while nodal voltage angles are bounded within $[-0.2,0.2]$~rad.
CIGRE benchmark nodal apparent-power values and associated power factors are used to construct the conventional active- and reactive-power demand profiles. These quantities represent coincident nodal demand and are not interpreted as local transformer capacities. CIGRE branch ampacities are represented through linear active-power branch limits derived from the corresponding current ratings. In addition, the nodal constraint in (15) uses a common active-power interface limit $\bar{P}^{\mathrm{grid}}_n=1000$~kW at the modeled charging locations. This parameter represents an adopted nodal hosting/interface-capacity limit and is distinct from both the conventional nodal demand and the branch thermal limits.}

\textcolor{black}{The framework incorporates a heterogeneous mix of single-port and multi-port hardware for both fast and slow charging}. Multi-port configurations are modeled as centralized units where a common power capacity is shared among four simultaneous connections, reflecting the hardware consolidation trends reported in \cite{nrel_costs_2023}. The cost structures and power ratings for these technologies are detailed in Table~\ref{tab:charger_specs}.
\begin{table}[h!]
\centering
\caption{Technical and economic parameters of the charging infrastructure.}
\label{tab:charger_specs}
\small
\begin{tabular}{llccc}
\toprule
\textbf{Type} & \textbf{Configuration} & \textbf{Power (kW)} & \textbf{Ports} & \textbf{Cost (€)} \\
\midrule
Slow & Single-Port & 7.5 & 1 & 1500 \\
Slow & Multi-Port (4$\times$) & 30 & 4 & 5,000 \\
Fast & Single-Port & 50 & 1 & 50,000 \\
Fast & Multi-Port (4$\times$) & 200 & 4 & 150,000 \\
\bottomrule
\end{tabular}
\end{table}

\subsection{EV fleet dynamics and battery modeling}\label{sec:EV_details}
The EV fleet size is varied from 250 to 600 vehicles to test scalability from moderate to high EV penetration in urban MV grids.
Spatial--temporal demand follows a deterministic home--work commuting pattern. Consumer nodes are split into residential and workplace sets, $\mathcal{N}^{\text{home}}$ and $\mathcal{N}^{\text{work}}$. Each EV is assigned $(n^{\text{home}}, n^{\text{work}})$, with locations and initial SOC sampled once to create a reproducible scenario. EVs are available at home at night (22:00--06:00) and at work during the day (08:00--18:00), shifting charging demand across time and space.

\textcolor{black}{The 40-kWh battery case is used as the reference configuration, while a 20-kWh case is evaluated as a battery-capacity sensitivity using the same fleet sizes and deterministic mobility realization. Charging efficiency is fixed at 85\%. To represent heterogeneity across EVs, each vehicle's initial SOC is sampled once from $\mathcal{U}(10\%,40\%)$, and the same realization is retained across the compared infrastructure configurations, based on reported mid-depletion charging behavior \cite{Corchero2014,HIPOLITO2022119065}.
The charging-service target is 100\% SOC. A minimum terminal SOC of 50\% is enforced during Stage~1 planning, whereas Stage~2A and Stage~2B use a 15\% terminal-SOC floor. Accordingly, the 100\% value in Stage~2 acts as a service target rather than a hard terminal requirement.}

\subsection{Operational validation and comparative setup}
\textcolor{black}{Configuration U is constructed from the Stage-1A solution O by preserving the total number of units of each charger technology while redistributing those units as evenly as possible across the eligible charging locations. Thus, for every $j\in\mathcal{J}$,
\begin{equation}
\sum_{n\in\mathcal{N}_c} N^{U}_{n,j}
=
\sum_{n\in\mathcal{N}_c} N^{O}_{n,j},
\qquad
\forall j\in\mathcal{J},
\label{eq:uniform_preservation}
\end{equation}
where $\mathcal{N}_c\subseteq\mathcal{N}$ denotes the eligible grid nodes. Hence, O and U have identical technology totals and infrastructure expenditure and differ only in spatial allocation.}

\textcolor{black}{Stage~2A evaluates O, I, and U with their infrastructure fixed and re-optimizes only the operational variables. The resulting final SOC, energy shortfall, and grid metrics therefore provide a common post-planning comparison.} 
\textcolor{black}{Stage~2B is reported for O and U at the 600-EV penetration level, allowing the effect of spatial redistribution to be examined while preserving the installed charger portfolio. 
For each reported Stage-2B case, the corresponding centralized Stage-2A schedule is retained as the baseline reference $p_{i,t}^{\mathrm{ref},k}$ prior to flexibility activation.
Thus, the reference is generated by the model rather than prescribed independently; operationally, it can be interpreted as the baseline or day-ahead charging schedule available before the DSO flexibility request.
The redispatched schedule $p_{i,t}^{M}$ is subsequently re-optimized with infrastructure fixed and the minimum 15\% terminal-SOC condition.}

\textcolor{black}{For the reported flexibility experiments, the allowable service degradation is set to
\begin{equation}
\Delta^{\mathrm{serv}}
\in
\left\{
0.01\,J_k^{\mathrm{serv},*},
\;
0.05\,J_k^{\mathrm{serv},*}
\right\},
\end{equation}
corresponding to the 1\% and 5\% cases reported in Table~\ref{tab:stage2b_final}. The DSO loading coefficient $\kappa$ is treated as an exogenous scenario parameter; a pre-dispatch screening of candidate $\kappa$ values is performed for the 600-EV cases, and $\kappa=0.425$ is adopted for the reported comparisons because it retains positive EV-addressable flexibility across both battery-capacity and infrastructure configurations. Accordingly, the same $\kappa$ is used as a common study-specific threshold for comparison and is not intended to represent a universal DSO operating limit.}

\section{Results and Discussion}\label{sec:results}
\textcolor{black}{The framework is implemented in Python with PuLP \cite{mitchell2011pulp}. Stage~1A/1B planning problems are solved with HiGHS~\cite{huangfu2018parallelizing}, while the fixed-infrastructure Stage~2A/2B problems use CBC \cite{forrest2005cbc}. 
A 5\% relative MIP-gap tolerance is adopted for all problems, and reported solutions satisfy the prescribed optimality gap. Where applicable, feasible planning solutions are supplied as MIP starts for Stage~1B; these warm starts affect only solver initialization and do not alter the formulation.
Runtime increases with fleet size (600-EV Stage~1: approximately 4~h), whereas the fixed-infrastructure operational problems in Stage~2 solve substantially faster. Results are reported for both 20- and 40-kWh battery capacities to assess sensitivity in Stage~1A planning and Stage~2A operational-validation. The 40-kWh case is used for the flexibility analyses.}

\subsection{Planning results and infrastructure scaling}

\textcolor{black}{
Tables~\ref{tab:planning_final_20kwh} and
\ref{tab:planning_final} summarize the Stage-1 infrastructure configurations for the 20-kWh and 40-kWh battery cases, respectively. Across both cases, the optimized portfolios contain only slow-charging technologies, with multi-port chargers supplying most of the available ports. Infrastructure requirements increase with fleet size and are consistently higher for the 40-kWh case: the number of available ports increases from 34 to 79 for the 20-kWh case and from 52 to 136 for the 40-kWh case as the fleet grows from 250 to 600 EVs. The number of physical charger units does not scale proportionally with the number of ports because the optimizer can substitute between single- and multi-port technologies.
}
\textcolor{black}{
For both battery capacities, configurations O and I have the same aggregate charger composition at 250, 350, 450, and 600 EVs, whereas different portfolios emerge at 550 EVs. At 550 EVs, in the 20-kWh case, O installs two slow single-port and 18 slow multi-port units (20 units and 74 ports), while I uses five single-port and 17 multi-port units (22 units and 73 ports). In the 40-kWh case, O installs eight slow single-port and 27 slow multi-port units (35 units and 116 ports), whereas I uses one single-port and 29 multi-port units (30 units and 117 ports). These results show that the Stage-1A investment limit does not uniquely determine the technology mix: the integrated reference can select a different combination of charger units and ports while remaining within the investment limit established by O.}

\textcolor{black}{Fig.~\ref{fig:dual_capex} compares Stage-1 CAPEX for configuration O in the 20-kWh and 40-kWh cases. CAPEX rises with fleet size in both, from 43 to 99.5~k\EUR (20 kWh) and from 66 to 170~k\EUR (40 kWh) as the fleet increases from 250 to 600 EVs. The 40-kWh case is costlier at all sizes, and the gap widens with fleet size, reflecting higher charging-energy needs and motivating the Stage-2A operational comparison.}

\begin{table}[htbp]
\centering
\caption{\textcolor{black}{Infrastructure configurations for the 20-kWh case.}}
\label{tab:planning_final_20kwh}

\setlength{\tabcolsep}{4.5pt}
\renewcommand{\arraystretch}{0.95}

\begin{tabular*}{\columnwidth}
{@{\extracolsep{\fill}} c|cc|cc|cc|cc@{}}
\hline

\multirow{2}{*}{Fleet}
& \multicolumn{2}{c|}{Slow single}
& \multicolumn{2}{c|}{Slow multi}
& \multicolumn{2}{c|}{Total units}
& \multicolumn{2}{c}{Total ports} \\

\cline{2-9}

& O & I
& O & I
& O & I
& O & I \\

\hline

250
& 2 & 2
& 8 & 8
& 10 & 10
& 34 & 34 \\

350
& 3 & 3
& 11 & 11
& 14 & 14
& 47 & 47 \\

450
& 1 & 1
& 15 & 15
& 16 & 16
& 61 & 61 \\

550
& 2 & 5
& 18 & 17
& 20 & 22
& 74 & 73 \\

600
& 3 & 3
& 19 & 19
& 22 & 22
& 79 & 79 \\

\hline
\end{tabular*}

\end{table}

\begin{table}[htbp]
\centering
\caption{\textcolor{black}{Infrastructure configurations for the 40-kWh case.}}
\label{tab:planning_final}

\setlength{\tabcolsep}{4.5pt}
\renewcommand{\arraystretch}{0.95}

\begin{tabular*}{\columnwidth}
{@{\extracolsep{\fill}} c|cc|cc|cc|cc@{}}
\hline

\multirow{2}{*}{Fleet}
& \multicolumn{2}{c|}{Slow single}
& \multicolumn{2}{c|}{Slow multi}
& \multicolumn{2}{c|}{Total units}
& \multicolumn{2}{c}{Total ports} \\

\cline{2-9}

& O & I
& O & I
& O & I
& O & I \\

\hline

250
& 4  & 4
& 12 & 12
& 16 & 16
& 52 & 52 \\

350
& 0  & 0
& 19 & 19
& 19 & 19
& 76 & 76 \\

450
& 5  & 5
& 23 & 23
& 28 & 28
& 97 & 97 \\

550
& 8  & 1
& 27 & 29
& 35 & 30
& 116 & 117 \\

600
& 0  & 0
& 34 & 34
& 34 & 34
& 136 & 136 \\

\hline
\end{tabular*}

\end{table}

\begin{figure}[htbp]
    \centering
    \input{capex_plot1.tex}
    \caption{\textcolor{black}{CAPEX for Stage-1 O configuration under 20-kWh and 40-kWh battery capacity cases across fleet sizes.}}
    \label{fig:dual_capex}
\end{figure}

\begin{table*}[htb!]
\centering
\caption{\textcolor{black}{Stage-2A centralized operational validation for the 20-kWh and 40-kWh EV cases.} 
}
\label{tab:stage2a_battery_comparison}

\setlength{\tabcolsep}{1.5pt}
\renewcommand{\arraystretch}{0.95}

\begin{tabular}{@{}c|JJJ|SSS|FFF|JJJ|SSS|FFF@{}}
\hline

\multirow{3}{*}{Fleet}
& \multicolumn{9}{c|}{20-kWh case}
& \multicolumn{9}{c}{40-kWh case} \\

\hhline{~|---------|---------}

&
\multicolumn{3}{c|}{$J^{\mathrm{serv}}$}
& \multicolumn{3}{c|}{Av. final SOC (\%)}
& \multicolumn{3}{c|}{Shortfall (kWh/EV)}
& \multicolumn{3}{c|}{$J^{\mathrm{serv}}$}
& \multicolumn{3}{c|}{Av. final SOC (\%)}
& \multicolumn{3}{c}{Shortfall (kWh/EV)} \\

\hhline{~|---|---|---|---|---|---}

& \textbf{O} & \textbf{I} & \textbf{U}
& \textbf{O} & \textbf{I} & \textbf{U}
& \textbf{O} & \textbf{I} & \textbf{U}
& \textbf{O} & \textbf{I} & \textbf{U}
& \textbf{O} & \textbf{I} & \textbf{U}
& \textbf{O} & \textbf{I} & \textbf{U} \\

\hline

250
& 11733.0 & 11918.8 & 12652.0
& 66.21 & 65.39 & 62.11
& 6.76 & 6.92 & 7.58
& 14109.7 & 14109.7 & 14773.2
& 57.92 & 57.92 & 54.99
& 16.83 & 16.83 & 18.00 \\

350
& 16279.2 & 16279.2 & 17588.6
& 66.95 & 66.95 & 62.80
& 6.61 & 6.61 & 7.44
& 19430.9 & 19430.9 & 20584.8
& 58.80 & 58.80 & 55.16
& 16.48 & 16.48 & 17.94 \\

450
& 21278.1 & 21278.1 & 22689.8
& 66.05 & 66.05 & 62.59
& 6.79 & 6.79 & 7.48
& 24948.7 & 24948.7 & 26387.6
& 58.97 & 58.97 & 55.44
& 16.41 & 16.41 & 17.83 \\

550
& 25843.6 & 26034.0 & 28046.4
& 66.45 & 66.30 & 61.98
& 6.71 & 6.74 & 7.60
& 30931.3 & 30726.4 & 32693.0
& 58.23 & 58.58 & 54.69
& 16.71 & 16.57 & 18.12 \\

600
& 28653.1 & 28653.1 & 30979.4
& 65.78 & 65.78 & 61.46
& 6.84 & 6.84 & 7.71
& 32138.3 & 32024.9 & 33978.0
& 60.62 & 60.83 & 57.23
& 15.75 & 15.67 & 17.11 \\

\hline
\end{tabular}

\end{table*}

\subsection{Operational validation and battery-capacity sensitivity}

\begin{figure}[htb]
\centering
\input{Figs/fig_avg_soc_600}
\caption{\textcolor{black}{Average SOC profile during Stage-2A operational validation for 600-EV, 40-kWh case.}}
\label{fig:avg_soc_600}
\end{figure}

\textcolor{black}{For the 600-EV, 40-kWh case in Fig.~\ref{fig:avg_soc_600}, configurations O and I have nearly identical average-SOC trajectories, with I ending slightly higher (60.62\% vs.\ 60.83\%). Configuration U follows a different pattern: its average SOC rises faster early but then lags, finishing at 57.23\%, indicating its poorer Stage-2A service as reported in Table~\ref{tab:stage2a_battery_comparison}.}

\textcolor{black}{Table~\ref{tab:stage2a_battery_comparison} compares O, I, and U under the Stage-2A objective. U consistently yields higher $J^{\mathrm{serv}}$ than O for both battery sizes: by about 6.6--8.5\% (20 kWh) and 4.7--5.9\% (40 kWh). Because U matches O in total chargers and CAPEX, the degradation is due to spatial redistribution, not reduced infrastructure.
Battery capacity strongly affects service: across fleet sizes, 20-kWh fleets achieve higher final SOC and lower energy shortfall than 40-kWh fleets. At 600 EVs, O reaches 65.78\% final SOC with a 6.84~kWh/EV shortfall (20 kWh) versus 60.62\% and 15.75~kWh/EV (40 kWh). The same O--U trend across all fleet sizes shows that the service impact of charger placement is not specific to one battery size.}

\textcolor{black}{The effect of the integrated reference I is small and case-dependent. For the 20-kWh cases, O and I are identical at 350, 450, and 600 EVs, while I gives slightly poorer service at 250 and 550 EVs. For the 40-kWh cases, the two configurations are identical at 250--450 EVs, whereas I provides modest improvements at 550 and 600 EVs; at 600 EVs, $J^{\mathrm{serv}}$ decreases from 32138.3 for O to 32024.9 for I. Thus, re-optimizing infrastructure within the Stage-1 CAPEX can improve post-planning service in some cases, but the benefit is neither large nor systematic.}

\begin{figure}[htb]
\centering
\input{Figs/fig_voltage_ou_600_v2}
\vspace{-0.8em}
\caption{\textcolor{black}{Hourly nodal-voltage distribution for the 600-EV, 40-kWh case under configurations O and U.}}
\label{fig:voltage_ou_600}
\end{figure}

\subsection{Spatial allocation, voltage profile, and local grid loading}

\textcolor{black}{The Stage-2A results show that spatial redistribution affects both charging-service quality and the temporal distribution of network stress. Fig.~\ref{fig:voltage_ou_600} compares the hourly nodal-voltage distributions for O and U. 
All nodal voltages remain within the imposed 0.95--1.05~p.u. operating range.}
\textcolor{black}{
Spatial redistribution does not uniformly improve the voltage profile. During the principal daytime charging interval, configuration U raises the lower-tail voltage relative to O, whereas the opposite behavior is observed during some early-day and late-hour periods. Both configurations exhibit the same overall minimum voltage, approximately at 0.97~p.u., for the reported 600-EV case. The result therefore indicates, the voltage stress shifts across nodes and time.
}

\textcolor{black}{A similar distinction applies to nodal loading. The maximum nodal load in the 600-EV case is dominated by the conventional demand profile and reaches approximately 705.7~kW for both O and U configurations. Hence, global peak loading alone does not capture the effect of charger placement.
The key issue is not the overall system peak, but rather the spatial and temporal alignment between controllable EV charging and locally strained node–time combinations, which motivates the Stage~2B flexibility analysis presented next.}

\begin{table}[t]
\centering
\caption{Stage-2B local-flexibility results for 600-EVs. 
}
\label{tab:stage2b_final}

\setlength{\tabcolsep}{1.35pt}
\renewcommand{\arraystretch}{0.95}
\scriptsize

\resizebox{\columnwidth}{!}{%
\begin{tabular}{cc|cc|cc|cc}
\hline

\multirow{4}{*}{\shortstack{Battery\\capacity}}
& \multirow{4}{*}{Config.}
& \multicolumn{2}{c|}{Pre-redispatch}
& \multicolumn{4}{c}{Service degradation} \\

\cline{3-8}

&
& \multirow{2}{*}{$E^{\mathrm{req}}$}
& \multirow{2}{*}{$E^{\mathrm{addr}}$}
& \multicolumn{2}{c|}{1\%}
& \multicolumn{2}{c}{5\%} \\

&
&
&
&
$E^{\mathrm{flex}}$
& $\rho^{\mathrm{addr}}$
& $E^{\mathrm{flex}}$
& $\rho^{\mathrm{addr}}$ \\

&
& (kWh)
& (kWh)
& (kWh)
& (\%)
& (kWh)
& (\%) \\

\hline

\multirow{2}{*}{20 kWh}
& O
& 1464.16 & 20.94
& 20.94 & 100.0
& 20.94 & 100.0 \\

& U
& 1517.63 & 74.40
& 72.17 & 97.0
& 74.40 & 100.0 \\

\hline

\multirow{2}{*}{40 kWh}
& O
& 2026.57 & 583.34
& 142.05 & 24.4
& 583.34 & 100.0 \\

& U
& 1611.37 & 168.15
& 157.61 & 93.7
& 168.15 & 100.0 \\

\hline
\end{tabular}%
}

\end{table}


\begin{figure}[t!]
\centering
\input{Figs/fig_market_flex_node8_600_1}
\vspace{-0.98em}
\caption{\textcolor{black}{Stage-2B DSO-requested and activated flexibility at Node~8 for 600-EVs, 40-kWh with configuration-O. 
}}
\label{fig:flex_node8_600}
\end{figure}


\subsection{Validation of local EV flexibility layer}

\textcolor{black}{
Table~\ref{tab:stage2b_final} summarizes the Stage-2B results for 600 EVs with $\kappa=0.425$. 
Since the DSO request reflects conventional and EV loading while only EV charging is controllable, $E^{\mathrm{req}}$ can exceed $E^{\mathrm{addr}}$; therefore, $\rho^{\mathrm{addr}}$ measures delivery relative to the EV-addressable quantity.}
\textcolor{black}{For the 20-kWh case, O activates its full 20.94~kWh of addressable flexibility at both service allowances, concentrated at Node~8. U activates 72.17~kWh of 74.40~kWh (97.0\%) at 1\% and the full amount at 5\%, distributed across Nodes~5 and~6. Thus, spatial redistribution changes both the amount and location of realizable flexibility.
The trade-off is stronger for the 40-kWh case. O activates only 142.05~kWh of 583.34~kWh (24.4\%) at 1\%, entirely at Node~8, but reaches full delivery at 5\%, extending to Nodes~8 and~10. 
U activates 157.61~kWh of 168.15~kWh (93.7\%) at 1\% and the full amount at 5\%, with flexibility delivered across Nodes~5,~6, and~8 under both allowances. Hence, both deployment options and service allowance shape the magnitude and spatial distribution of flexibility.}

\textcolor{black}{Fig.~\ref{fig:flex_node8_600} illustrates this behavior at Node~8 for the 600-EVs, 40-kWh configuration-O. Under the 1\% allowance, only part of the requested flexibility can be activated while maintaining the prescribed charging-service quality. With a 5\% allowance, the aggregator can shift a larger amount of charging, causing the post-flexibility nodal load to follow the DSO loading threshold more closely over the requested period. At some hours, however, a DSO request is present while little or no reference EV charging is available for reduction. This mismatch explains why the total request $E^{\mathrm{req}}$ can exceed the EV-addressable quantity $E^{\mathrm{addr}}$.}
\textcolor{black}{A uniform illustrative remuneration of
$\lambda^{\mathrm{flex}}=0.25$~\EUR/kWh is applied to all requested flexibility intervals across the considered nodes. Given the strictly positive and uniform remuneration, the monetary value of accepted flexibility is scaled without changing the relative preference among feasible redispatch decisions.}

\section{Conclusions}
\textcolor{black}{For large EV fleets, this work develops a sequential planning--operation framework to evaluate deployed charging infrastructure under distribution-grid operational constraints. 
By comparing deployment options O, U, and I under a common operational criterion, the framework isolates the impacts of investment, siting, and operational EV charging flexibility.
Results show that infrastructure requirements increase with EV fleet size and battery capacity, while similar aggregate charging capacity can yield different service levels.
Redistributing the hardware in U consistently degrades the Stage-2A centralized service objective relative to O, although the resulting voltage and loading effects vary across time and location. 
The reference I produces only small, case-dependent changes, with modest improvements in some 40-kWh cases.}

\textcolor{black}{Stage~2B further shows that DSO flexibility requests exceed the EV-addressable portion of load. With a 1\% service-degradation allowance, EV-addressable delivery ranges from 24.4\% to 100\% across the reported cases, whereas a 5\% allowance permits full delivery of the EV-addressable quantity. 
Realizable flexibility therefore depends on the deployment configuration, the temporal and spatial coincidence of controllable EV charging with local grid stress, and the service degradation permitted during redispatch.
Future work will incorporate additional network operating conditions and more detailed local-flexibility coordination mechanisms.}

\balance

\bibliographystyle{IEEEtran}
\bibliography{./references.bib}

\end{document}

%% file: capex_plot1.tex
\begin{tikzpicture}
\begin{axis}[
    ybar=2pt,
    width=0.99\columnwidth,
    height=4.2cm,
    bar width=9pt,
    axis lines=box,
    ymin=0,
    ymax=185,
    ytick={0,50,100,150},
    ylabel={Investment cost (k\EUR)},
    xlabel={Number of EVs in fleet},
    symbolic x coords={250,350,450,550,600},
    xtick=data,
    enlarge x limits=0.12,
    ymajorgrids=false,
    xmajorgrids=false,
    tick label style={font=\footnotesize},
    label style={font=\footnotesize},
    legend style={
        font=\scriptsize,
        at={(0.03,0.97)},
        anchor=north west,
        legend columns=1,
        draw=black,
        fill=white,
        inner xsep=3pt,
        inner ysep=2pt,
        row sep=1pt
    }
]

\addplot[fill=blue!70!black, draw=black] coordinates {
    (250,43.0)
    (350,59.5)
    (450,76.5)
    (550,93.0)
    (600,99.5)
};

\addplot[fill=red!70!black, draw=black] coordinates {
    (250,66.0)
    (350,95.0)
    (450,122.5)
    (550,147.0)
    (600,170.0)
};

\legend{20-kWh case, 40-kWh case}

\end{axis}
\end{tikzpicture}

%% file: Figs/fig_avg_soc_600.tex
\pgfplotstableread[col sep=space]{
hour O I U
0 24.9327 24.9327 24.9327
1 25.1452 25.0389 26.8452
2 25.3577 25.1452 28.7577
3 25.5702 25.2514 30.6702
4 25.7827 25.3577 32.5827
5 25.9952 25.4639 34.4952
6 26.2077 25.5702 36.4077
7 26.2077 25.5702 36.4077
8 26.2077 25.5702 36.4077
9 29.6077 29.0764 38.1077
10 33.0077 32.5827 39.8077
11 36.4077 36.0889 41.5077
12 39.8077 39.5952 43.2077
13 43.2077 43.1014 44.9077
14 46.6077 46.6077 46.6077
15 50.0077 50.1139 48.3056
16 53.4077 53.6196 50.0056
17 56.8077 57.1247 51.7056
18 60.1946 60.6165 53.4056
19 60.1946 60.6165 53.4056
20 60.1946 60.6165 53.4056
21 60.1946 60.6165 53.4056
22 60.1946 60.6165 53.4056
23 60.4071 60.7227 55.3181
24 60.6196 60.8290 57.2298
}\socTable

\begin{tikzpicture}
\begin{axis}[
width=\columnwidth,
height=3.8cm,
xmin=0, xmax=23,
xtick={0,6,12,18,23},
ymin=20, ymax=62,
ytick={20,30,40,50,60},
grid=both,
grid style={draw=gray!20, line width=.1pt},
major grid style={draw=gray!35, line width=.2pt},
xlabel={Hour index},
ylabel={Av. fleet SOC (\%)},
tick label style={font=\footnotesize},
label style={font=\footnotesize},
legend style={
    font=\scriptsize,
    at={(0.03,0.97)},
    anchor=north west,
    legend columns=1,
    draw=black,
    fill=white,
    cells={anchor=west},
    row sep=1pt,
    inner xsep=3pt,
    inner ysep=2pt
}
]
\addplot[blue, thick, mark=*, mark size=1.3pt] 
table[x=hour,y=O]{\socTable};

\addplot[red!75!black, dashed, thick, mark=square*, mark size=1.3pt] 
table[x=hour,y=I]{\socTable};

\addplot[black, dotted, thick, mark=triangle*, mark size=1.5pt] 
table[x=hour,y=U]{\socTable};

\legend{O,I,U}
\end{axis}
\end{tikzpicture}

%% file: Figs/fig_voltage_ou_600_v2.tex

\pgfplotstableread[col sep=space]{
hour p25 p50 p75 pmin
0 0.989649 0.989994 0.999370 0.989395
1 0.991076 0.991379 0.999460 0.990861
2 0.991864 0.992144 0.999510 0.991670
3 0.992696 0.992952 0.999562 0.992526
4 0.993231 0.993472 0.999596 0.993076
5 0.993015 0.993262 0.999583 0.992854
6 0.992214 0.992440 0.999508 0.991998
7 0.989609 0.989910 0.999343 0.989321
8 0.981869 0.982890 0.996784 0.981068
9 0.980975 0.982019 0.996727 0.980145
10 0.979955 0.981024 0.996662 0.979092
11 0.978880 0.979976 0.996594 0.977983
12 0.978236 0.979348 0.996553 0.977319
13 0.976730 0.977880 0.996458 0.975764
14 0.975750 0.976924 0.996395 0.974753
15 0.976134 0.977299 0.996420 0.975150
16 0.977606 0.978733 0.996513 0.976668
17 0.977177 0.978312 0.996484 0.976225
18 0.977559 0.978210 0.998581 0.976937
19 0.971597 0.972422 0.998204 0.970810
20 0.969106 0.970003 0.998046 0.968250
21 0.972811 0.973600 0.998280 0.972057
22 0.977568 0.978263 0.998606 0.976978
23 0.984429 0.984925 0.999040 0.984030
}\voltageOstat

\pgfplotstableread[col sep=space]{
hour p25 p50 p75 pmin
0 0.986203 0.986529 0.999370 0.985822
1 0.987622 0.987914 0.999460 0.987279
2 0.988406 0.988679 0.999510 0.988083
3 0.989234 0.989487 0.999562 0.988934
4 0.989767 0.990007 0.999596 0.989480
5 0.989551 0.989797 0.999583 0.989260
6 0.992214 0.992440 0.999508 0.991998
7 0.989609 0.989910 0.999343 0.989321
8 0.984168 0.984888 0.998325 0.983540
9 0.983274 0.984016 0.998268 0.982618
10 0.982253 0.983021 0.998203 0.981565
11 0.981178 0.981974 0.998135 0.980456
12 0.980526 0.981346 0.998094 0.979791
13 0.979000 0.979877 0.997999 0.978237
14 0.978013 0.978926 0.997936 0.977231
15 0.978397 0.979296 0.997961 0.977622
16 0.979887 0.980731 0.998054 0.979141
17 0.979421 0.980283 0.998025 0.978666
18 0.977559 0.978210 0.998581 0.976937
19 0.971597 0.972422 0.998204 0.970810
20 0.969106 0.970003 0.998046 0.968250
21 0.972811 0.973600 0.998280 0.972057
22 0.974134 0.974798 0.998606 0.973483
23 0.980996 0.981461 0.999040 0.980491
}\voltageUstat

\begin{tikzpicture}

\begin{groupplot}[
group style={
    group size=1 by 2,
    vertical sep=0.45cm
},
width=\columnwidth,
height=3.1cm,
xmin=0, xmax=23,
xtick={0,6,12,18,23},
ymin=0.968, ymax=1.001,
ytick={0.970,0.980,0.990,1.000},
yticklabels={0.970,0.980,0.990,1.000},
scaled y ticks=false,
grid=both,
grid style={draw=gray!20, line width=.1pt},
major grid style={draw=gray!35, line width=.2pt},
tick label style={font=\footnotesize},
label style={font=\footnotesize},
every axis plot/.append style={line join=round}
]

\nextgroupplot[
ylabel={V mag. (p.u.)},
xticklabels=\empty
]

\addplot[name path=upperO, draw=none, forget plot]
table[x=hour,y=p75]{\voltageOstat};

\addplot[name path=lowerO, draw=none, forget plot]
table[x=hour,y=p25]{\voltageOstat};

\addplot[blue!25, forget plot]
fill between[of=upperO and lowerO];

\addplot[blue!85!black, line width=1.0pt, forget plot]
table[x=hour,y=p50]{\voltageOstat};

\addplot[green!45!black, dashed, line width=1.25pt, forget plot]
table[x=hour,y=pmin]{\voltageOstat};

\addplot[blue!45!black, dashed, line width=1.0pt, forget plot]
coordinates {(0,1.0) (23,1.0)};

\node[anchor=west, font=\footnotesize] at (axis cs:1.0,0.9740)
{Configuration O};

\nextgroupplot[
xlabel={Hour},
ylabel={V mag. (p.u.)},
legend style={
    font=\scriptsize,
    at={(0.5,-0.56)},
    anchor=north,
    legend columns=4,
    draw=black,
    fill=white,
    fill opacity=1,
    text opacity=1,
    inner xsep=5pt,
    inner ysep=4pt,
    /tikz/every even column/.append style={column sep=10pt}
}
]

\addplot[name path=upperU, draw=none, forget plot]
table[x=hour,y=p75]{\voltageUstat};

\addplot[name path=lowerU, draw=none, forget plot]
table[x=hour,y=p25]{\voltageUstat};

\addplot[blue!25, forget plot]
fill between[of=upperU and lowerU];

\addplot[blue!85!black, line width=1.0pt, forget plot]
table[x=hour,y=p50]{\voltageUstat};

\addplot[green!45!black, dashed, line width=1.25pt, forget plot]
table[x=hour,y=pmin]{\voltageUstat};

\addplot[blue!45!black, dashed, line width=1.25pt, forget plot]
coordinates {(0,1.0) (23,1.0)};

\node[anchor=west, font=\footnotesize] at (axis cs:1.0,0.9740)
{Configuration U};

\addlegendimage{area legend, draw=blue!85!black, fill=blue!25}
\addlegendentry{25--75\%}

\addlegendimage{blue!85!black, line width=1.2pt}
\addlegendentry{Median}

\addlegendimage{green!45!black, dashed, line width=1.4pt}
\addlegendentry{Min.}

\addlegendimage{blue!45!black, dashed, line width=1.2pt}
\addlegendentry{Slack}

\end{groupplot}

\end{tikzpicture}

%% file: Figs/fig_market_flex_node8_600_1.tex


\pgfplotstableread[col sep=space]{
hour ref post request activated
0 183.632 183.632 0.000 0.000
1 157.350 157.350 0.000 0.000
2 142.833 142.833 0.000 0.000
3 127.495 127.495 0.000 0.000
4 117.629 117.629 0.000 0.000
5 121.613 121.613 0.000 0.000
6 143.476 143.476 0.000 0.000
7 191.483 191.483 0.000 0.000
8 416.256 416.256 0.000 0.000
9 432.797 432.797 7.797 0.000
10 451.676 451.676 26.676 0.000
11 471.560 471.560 46.560 0.000
12 483.477 483.477 58.477 0.000
13 511.349 511.349 86.349 0.000
14 529.476 529.476 104.476 0.000
15 522.373 522.373 97.373 0.000
16 495.144 428.249 70.144 66.895
17 500.152 425.000 75.152 75.152
18 413.528 413.528 0.000 0.000
19 523.375 523.375 98.375 0.000
20 569.290 569.290 144.290 0.000
21 501.019 501.019 76.019 0.000
22 406.263 406.263 0.000 0.000
23 279.832 279.832 0.000 0.000
}\flexTableOne

\pgfplotstableread[col sep=space]{
hour ref post request activated
0 183.632 183.632 0.000 0.000
1 157.350 157.350 0.000 0.000
2 142.833 142.833 0.000 0.000
3 127.495 127.495 0.000 0.000
4 117.629 117.629 0.000 0.000
5 121.613 121.613 0.000 0.000
6 143.476 143.476 0.000 0.000
7 191.483 191.483 0.000 0.000
8 416.256 416.256 0.000 0.000
9 432.797 425.000 7.797 7.797
10 451.676 425.000 26.676 26.676
11 471.560 425.000 46.560 46.560
12 483.477 425.000 58.477 58.477
13 511.349 425.000 86.349 86.349
14 529.476 425.000 104.476 104.476
15 522.373 425.000 97.373 97.373
16 495.144 425.000 70.144 70.144
17 500.152 425.000 75.152 75.152
18 413.528 413.528 0.000 0.000
19 523.375 523.375 98.375 0.000
20 569.290 569.290 144.290 0.000
21 501.019 501.019 76.019 0.000
22 406.263 406.263 0.000 0.000
23 279.832 279.832 0.000 0.000
}\flexTableFive

\begin{tikzpicture}

\begin{axis}[
name=topLeft,
width=0.9\columnwidth,
height=3.0cm,
xmin=0, xmax=23,
xtick={0,6,12,18,23},
xticklabels=\empty,
ymin=0, ymax=620,
ytick={0,200,400,600},
ylabel={N8 load (kW)},
grid=both,
grid style={draw=gray!20, line width=.1pt},
major grid style={draw=gray!35, line width=.2pt},
tick label style={font=\footnotesize},
label style={font=\footnotesize},
title style={font=\footnotesize},
title={(a) 1\% service-degradation allowance},
axis lines=box,
legend to name=flexlegend,
legend columns=3,
legend style={
    font=\scriptsize,
    draw=black,
    fill=white,
    column sep=4pt
}
]
\addplot[blue, thick] table[x=hour,y=ref]{\flexTableOne};
\addlegendentry{Reference}

\addplot[red!75!black, dashed, thick] table[x=hour,y=post]{\flexTableOne};
\addlegendentry{Post-flex.}

\addplot[gray!65, densely dashdotted, thick] coordinates {(0,425) (23,425)};
\addlegendentry{Trigger ($\kappa\bar P_{n8}^{\mathrm{grid}}$)}

\addlegendimage{green!60!black, thick, mark=square*, mark size=1.3pt}
\addlegendentry{Request ($R_{n8,t}^{\mathrm{flex}}$)}

\addlegendimage{black, dotted, thick, mark=*, mark size=1.2pt}
\addlegendentry{Activated ($f_{n8,t}^{\mathrm{flex}}$)}
\end{axis}

\begin{axis}[
at={(topLeft.south west)},
anchor=south west,
width=0.9\columnwidth,
height=3.0cm,
xmin=0, xmax=23,
xtick=\empty,
ymin=0, ymax=160,
ytick={0,50,100,160},
ylabel={Req./act. flex (kW)},
tick label style={font=\footnotesize},
label style={font=\footnotesize},
axis x line=none,
axis y line*=right,
ymajorgrids=false
]
\addplot[
green!60!black,
thick,
mark=square*,
mark size=1.3pt
]
table[x=hour,y=request]{\flexTableOne};

\addplot[
black,
dotted,
thick,
mark=*,
mark size=1.2pt
]
table[x=hour,y=activated]{\flexTableOne};
\end{axis}

\begin{axis}[
name=bottomLeft,
at={(topLeft.south west)},
anchor=north west,
yshift=-1.15cm,
width=0.9\columnwidth,
height=3.0cm,
xmin=0, xmax=23,
xtick={0,6,12,18,23},
ymin=0, ymax=620,
ytick={0,200,400,600},
xlabel={Hour},
ylabel={N8 load (kW)},
grid=both,
grid style={draw=gray!20, line width=.1pt},
major grid style={draw=gray!35, line width=.2pt},
tick label style={font=\footnotesize},
label style={font=\footnotesize},
title style={font=\footnotesize},
title={(b) 5\% service-degradation allowance},
axis lines=box
]
\addplot[blue, thick] table[x=hour,y=ref]{\flexTableFive};

\addplot[red!75!black, dashed, thick] table[x=hour,y=post]{\flexTableFive};

\addplot[gray!65, densely dashdotted, thick] coordinates {(0,425) (23,425)};
\end{axis}

\begin{axis}[
at={(bottomLeft.south west)},
anchor=south west,
width=0.9\columnwidth,
height=3.0cm,
xmin=0, xmax=23,
xtick=\empty,
ymin=0, ymax=160,
ytick={0,50,100,160},
ylabel={Req./act. flex (kW)},
tick label style={font=\footnotesize},
label style={font=\footnotesize},
axis x line=none,
axis y line*=right,
ymajorgrids=false
]
\addplot[
green!60!black,
thick,
mark=square*,
mark size=1.3pt
]
table[x=hour,y=request]{\flexTableFive};

\addplot[
black,
dotted,
thick,
mark=*,
mark size=1.2pt
]
table[x=hour,y=activated]{\flexTableFive};
\end{axis}

\node[
anchor=north,
font=\scriptsize
] at ($(bottomLeft.south)+(0,-1.0cm)$) {\pgfplotslegendfromname{flexlegend}};

\end{tikzpicture}